\documentclass[%
 aip,
 apl,%
 amsmath,amssymb,
 nopreprintnumbers,
 floatfix,
 reprint,%
]{revtex4-1}

\usepackage{graphicx}
\usepackage{dcolumn}
\usepackage{bm}

\newcommand{\Deg}{\ensuremath{^\circ}} 
\newcommand{\DegC}{\ensuremath{^\circ}C} 

\begin{document}

\preprint{AIP/123-QED}

\title[Fe/Ba-122 Bilayers]{Coherent interfacial bonding on the FeAs tetrahedron in Fe/Ba(Fe$_{1-x}$Co$_x$)$_2$As$_2$ bilayers}

\author{T. Therslef\mbox{}f}\email{T.Thersleff@ifw-dresden.de.}
 \affiliation{IFW Dresden, Institute for Metallic Materials, P.O. Box 270116, D-01171 Dresden, Germany}
\author{K. Iida}\email{K.Iida@ifw-dresden.de.}
 \affiliation{IFW Dresden, Institute for Metallic Materials, P.O. Box 270116, D-01171 Dresden, Germany}
\author{S. Haindl} \affiliation{IFW Dresden, Institute for Metallic Materials, P.O. Box 270116, D-01171 Dresden, Germany}
\author{M. Kidszun} \affiliation{IFW Dresden, Institute for Metallic Materials, P.O. Box 270116, D-01171 Dresden, Germany}
\author{D. Pohl} \affiliation{IFW Dresden, Institute for Metallic Materials, P.O. Box 270116, D-01171 Dresden, Germany}
\author{A. Hartmann} \affiliation{IFW Dresden, Institute for Metallic Materials, P.O. Box 270116, D-01171 Dresden, Germany}
\author{F. Kurth} \affiliation{IFW Dresden, Institute for Metallic Materials, P.O. Box 270116, D-01171 Dresden, Germany}
\author{J. H{\"a}nisch} \affiliation{IFW Dresden, Institute for Metallic Materials, P.O. Box 270116, D-01171 Dresden, Germany}
\author{R. H{\"u}hne} \affiliation{IFW Dresden, Institute for Metallic Materials, P.O. Box 270116, D-01171 Dresden, Germany}
\author{B. Rellinghaus} \affiliation{IFW Dresden, Institute for Metallic Materials, P.O. Box 270116, D-01171 Dresden, Germany}
\author{L. Schultz} \affiliation{IFW Dresden, Institute for Metallic Materials, P.O. Box 270116, D-01171 Dresden, Germany}
\author{B. Holzapfel} \affiliation{IFW Dresden, Institute for Metallic Materials, P.O. Box 270116, D-01171 Dresden, Germany}

\date{\today}

\begin{abstract}

We demonstrate the growth of epitaxial Fe/Ba(Fe$_{1-x}$Co$_x$)$_2$As$_2$ (Fe/Ba--122) bilayers on MgO(001) and LSAT(001) single crystal substrates using Pulsed Laser Deposition (PLD). By exploiting the metallic nature of the FeAs tetrahedron in the Ba-122 crystal structure, we achieve a coherent interfacial bond between bcc iron and Co-doped Ba--122. $T_{\rm c}$ values for both bilayers were close to that of the PLD target. Direct observation of interfacial bonding between Fe and the Ba--122 FeAs sublattice by atomic resolution transmission electron microscopy implies that this bilayer architecture may work for other iron pnictide systems and pave the way for the fabrication of superconducting/ferromagnetic heterostructures.

\end{abstract}

\maketitle

Rapidly following the discovery of superconductivity in the AE(Fe$_{1-x}$Co$_x$)$_2$As$_2$ (AE-122, AE = Alkaline Earth) system\cite{Rotter2008}, thin films were produced to probe the fundamental properties of and assess potential applications for these unique materials\cite{Hiramatsu2008,Choi2009,Hiramatsu2009a,Katase2009a,Lee2009,Iida2009,Lee2010,Iida2010a,Katase2010}.  However, difficulties overcoming the poor metal/oxide bond at the interface of many substrates has necessitated the need for significant optimization of the deposition parameters\cite{Iida2010a,Katase2010} as well as the use of various intermediate layers\cite{Lee2010} to produce well-textured films.  In spite of these efforts, nearly all of these films contain an unintentional amorphous or iron-containing layer at the interface.  While the nature of this interface is not yet fully understood, the disruption of local crystallographic ordering associated with it precludes the use of these films for interface-sensitive applications such as multilayers or heterostructures where coherent and chemically inert phase boundaries are required.  Moreover, it may be responsible for the challenging growth of epitaxial Ba--122 films in general\cite{Iida2009,Iida2010a} as well as the generation of pinning-active columnar defects observed to originate at this interface in some films\cite{Lee2010}.

\begin{figure}[b]
	\centering
		\includegraphics{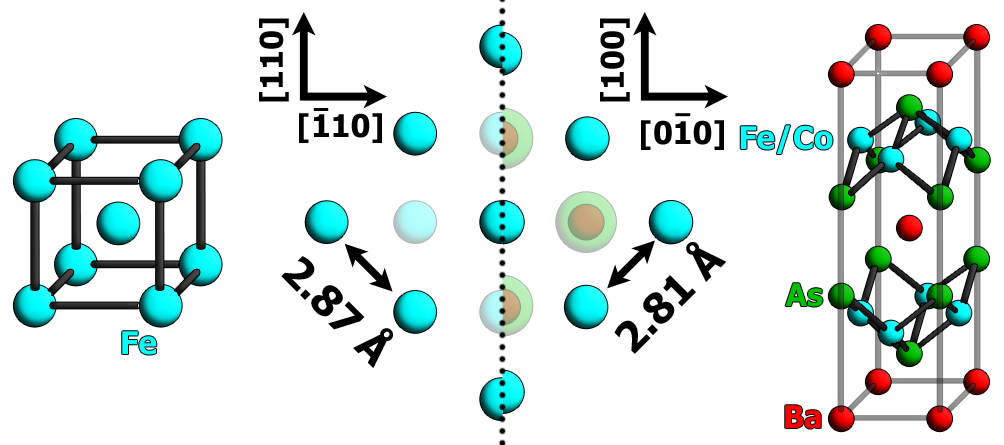}
		\caption{(Color online) The mismatch between the (001) surface plane of bcc iron (left) and the iron sublayer in the Co-doped Ba--122 unit cell (right) is about 2\%\ . Atomic radii are not to scale.}
	\label{fig:Model}
\end{figure}

A careful TEM investigation of the interface between epitaxially-grown Co-doped Ba--122 and bare (La,Sr)(Al,Ta)O$_3$ (LSAT) substrates revealed significant amounts of textured body-centered cubic (bcc) iron\cite{Iida2010a}.  The orientation of this iron layer was rotated 45\Deg\ in-plane to both the substrate and the Ba--122 phase and is the likely culprit for the Fe (200) reflection in XRD patterns.  In this orientation, the (100) surface plane of iron has an approximately 2\%\ lattice mismatch with the square-planar iron sublayer defining the FeAs tetrahedron in the Ba--122 unit cell and thus offers a natural location for metallic bonding (figure \ref{fig:Model}).  Furthermore, since the quality of our films containing these iron regions is consistently very high, it appears to be advantageous to their epitaxial growth and superconducting properties.

To investigate the nature of bonding at this interface, we deposited Fe/Ba--122 bilayers on MgO(001) and LSAT(001) substrates.  An Fe layer of 20 nm was deposited onto MgO and LSAT at 620\DegC\ using the standard on-axis Pulsed Laser Deposition (PLD) technique in a $10^{-8}$ mbar chamber with a 248 nm KrF laser operating at 10Hz.  Subsequently, a Ba--122 layer of around 130 nm was deposited at 700\DegC .  We observed island growth of Fe in the resulting films on both substrates; however, considerable research on the optimization of iron thin film growth on MgO already exists\cite{Jordan1998}.  Accordingly, iron on the MgO substrate was first deposited at room temperature and then heated to 700\DegC\ for the deposition of the Ba--122 phase under identical conditions used for the Ba--122 layer on Fe-buffered LSAT.

\begin{figure}
	\centering
		\includegraphics[width=8.5cm]{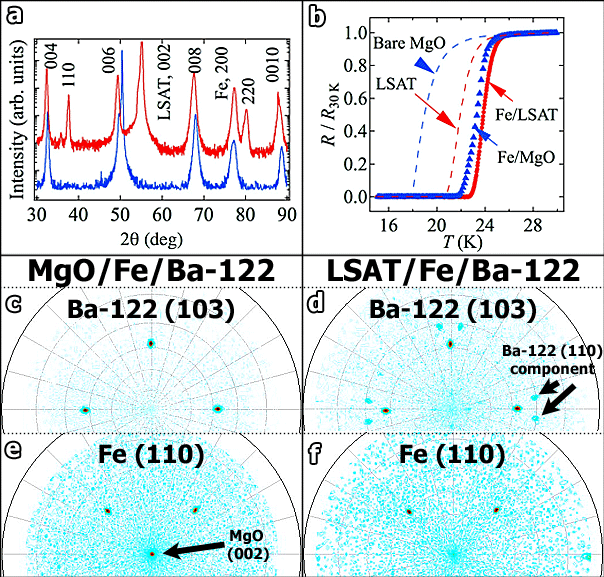}
		\caption{(Color online) (a) X-Ray diffraction pattern for bilayers on both LSAT and MgO in the Bragg-Brentano geometry.  Neither bilayer shows evidence for secondary phases and both exhibit a strong \textit{c}-axis texture. The LSAT bilayer additionally contains a second texture component evidenced by the Ba--122 (110) reflection.  (b)  Resistively measured $T_{\rm c}$ curves for Ba--122 deposited upon bare LSAT and MgO as well as the new Fe/Ba--122 bilayers.  $T_{\rm c,90}$ is 24.4 K and 24.8 K for the MgO and LSAT bilayers respectively. (c)  Pole figure of the Ba--122 (103) reflection for the  MgO bilayer reveals highly textured growth with a $\Delta \phi$ value of 0.95\Deg . (d) While the Ba--122 (103) reflection exhibits excellent texture with a $\Delta \phi$ value of 1.17\Deg , the additional (110) component also appears to be textured. (e,f) Iron on both substrates grows with a 45\Deg\ in-plane rotation and exhibits four-fold symmetry.  The additional peak in the center of (e) arises from the (002) plane of the MgO substrate.  All pole figures are plotted on a square root scale.}
	\label{fig:XRD}
\end{figure}

The XRD data acquired using the Bragg-Brentano geometry with Co $K_{\alpha}$ radiation (figure \ref{fig:XRD}a) reveal \textit{c}-axis textured growth for both iron and Ba--122 layers on MgO and LSAT substrates.  Neither bilayer shows evidence for secondary phases. The additional Ba--122 texture component on the LSAT substrate leading to the (110) peak in the XRD scan has a distinct epitaxial relationship to the substrate with $(110)[001]$Ba--122 $\parallel$ $(001)[110]$ LSAT and $(110)[001]$Ba--122 $\parallel$ $(001)[\overline{1}10]$ LSAT. As a result, two additional satellite peaks appear in the (103) pole figure near the peak for the main texture component (figure \ref{fig:XRD}d).  On MgO, no additional texture components could be identified suggesting pure epitaxial growth (figure \ref{fig:XRD}c).  The Ba--122 (103) reflection on both substrates exhibits four-fold symmetry with $\Delta \phi _{\textup{MgO,LSAT}} = 0.95^{\circ},1.17^{\circ}$ and (004) rocking curves reveal $\Delta \omega _{\textup{MgO,LSAT}} = 0.64^{\circ},1.01^{\circ}$.  In figure \ref{fig:XRD}e,f, the iron layer appears well textured on both substrates with a 45\Deg\ in-plane rotation to the Ba--122 phase.

Resistively-measured $T_{\rm c}$ values (figure \ref{fig:XRD}b) for both films are very high showing a $T_{c,90}$ of 24.4 K and 24.8 K on MgO and LSAT respectively. These values are among the highest reported for any Co-doped Ba--122 thin film to date and are nearly equal to that of the PLD target used ($T_{\rm c} = 25.5$ K as measured with a vibrating sample magnetometer).

\begin{figure}
	\centering
		\includegraphics{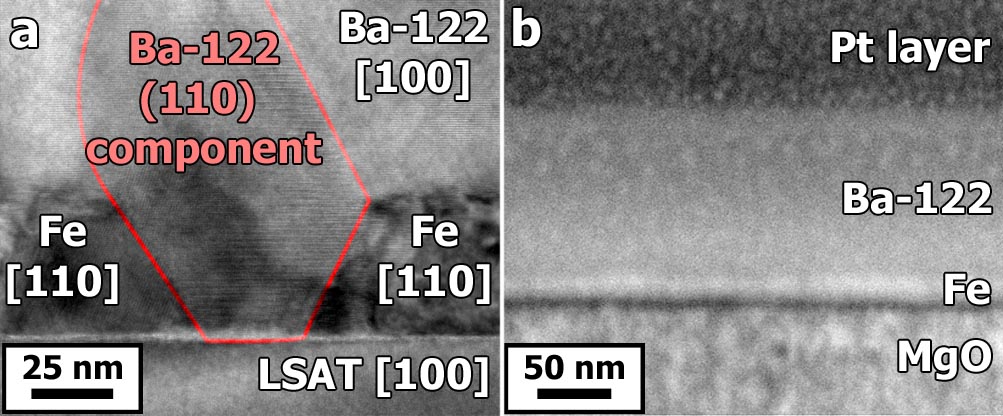}
		\caption{(Color online) (a) Bright field TEM overview of two faceted iron islands. Ba--122 between iron islands grows with $(110)[001]$Ba--122 $\parallel$ $(001)[110]$ LSAT and $(110)[001]$Ba--122 $\parallel$ $(001)[\overline{1}10]$ LSAT whereas atop the (100) surface facet, Ba--122 grows with $(001)[100]$ Fe $\parallel (001)[110]$ Ba--122.  (b) FIB cut imaged using a secondary electron in-lens detector on the MgO bilayer reveals no island growth.}
	\label{fig:Figure2}
\end{figure}

To reveal the location of the misaligned Ba--122 on the LSAT bilayer as well as to elucidate the nature of the Fe/Ba--122 interface common to the bilayers on both substrates, a comprehensive Transmission Electron Microscopy (TEM) investigation on this film was initiated.  A TEM lamella was prepared using the Focused Ion Beam (FIB) in-situ lift-out technique\cite{Langford06}.  The bright field TEM overview shown in figure \ref{fig:Figure2}a confirms the nucleation of the (001) faceted iron islands discussed previously.  Significantly, the (110) oriented Ba--122 component observed in figure \ref{fig:XRD}a,d appears to grow exclusively between these faceted islands whereas the (001) iron surface plane provides an effective interface for the epitaxial growth of the Ba--122 phase, showing no misalignment of the texture over large sample areas.  On the MgO substrate, the optimized deposition parameters for the iron layer eliminated any island growth as evidenced by a scanning electron microscopy image of a FIB cross-section provided in figure \ref{fig:Figure2}b.  Consequently, no misaligned texture is observed in figure \ref{fig:XRD} for the MgO bilayer.

\begin{figure}
	\centering
		\includegraphics{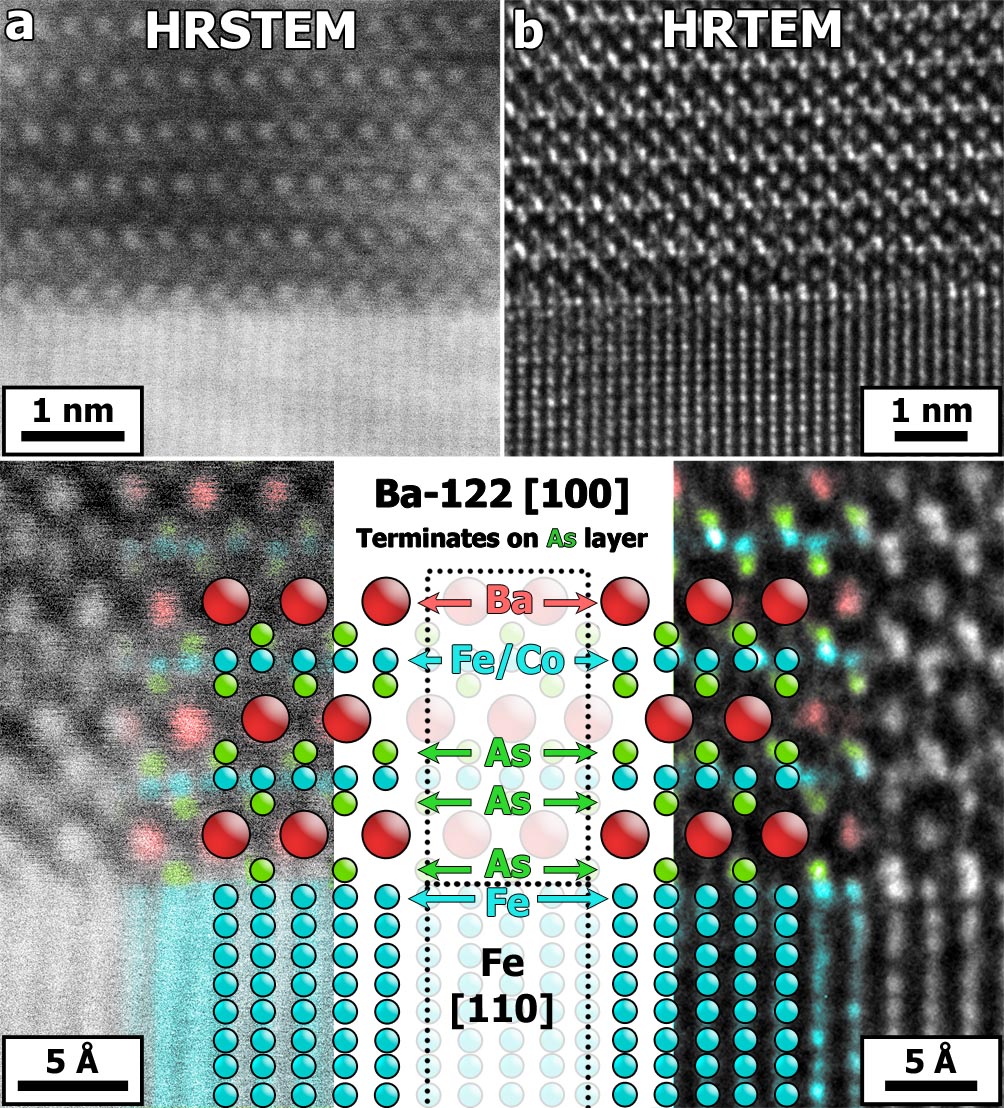}
		\caption{(Color online) (a) HRSTEM image of the Fe/Ba--122 interface.  Directly below, a higher resolution image of a neighboring region is shown.  (b) HRTEM image from the Fe/Ba--122 interface with an enlargement below. In both (a) and (b), the atomic columns appear white and are identified through the use of artificial coloration and a schematic model.  The interface is highly coherent and bonding takes place directly on the FeAs sublattice.}
	\label{fig:Figure3}
\end{figure}

The interface between Fe and Ba--122 was studied on the TEM lamella described above using High Resolution Scanning TEM (HRSTEM) on an FEI Titan$^3$ 80--300 microscope with an image $C_s$ corrector operating at 300 kV.  Figure \ref{fig:Figure3}a shows a HRSTEM image obtained with a High Angle Annular Dark Field (HAADF) detector.  Directly below, higher resolution data obtained from a different sample region is presented.  By selecting a camera length of 363 mm, the atomic columns appear as bright dots and the crystallographic symmetry of the Ba--122 phase becomes evident.  In the lower portion of figure \ref{fig:Figure3}, the location of atomic columns is denoted using artificial colors and a schematic model.  The Ba--122 phase is observed to terminate on the upper As sublayer of the FeAs tetrahedron.

As an independent confirmation of this analysis, High Resolution TEM (HRTEM) was undertaken on a separate sample region and is presented in figure \ref{fig:Figure3}b.  This image was acquired such that the atomic columns appear as white dots, significantly easing the image interpretation.  Below, the atomic positions are identified in the same manner as previously.  Combined with the HRSTEM data in figure \ref{fig:Figure3}a, these observations constitute compelling evidence that the square-planar iron sublayer in the Ba--122 unit cell is directly replaced by the (001) surface plane of the bcc iron layer resulting in a coherent interfacial bond on the FeAs sublattice.


The results of this study contain some wide-reaching implications.  First, the metallic nature of and excellent lattice matching between the Ba--122 iron sublayer and the bcc iron (001) surface plane ensure that this interface will be highly coherent.  This suggests that the epitaxial growth of Ba--122 will be favorable on any substrate upon which a planar iron (001) facet can be grown, as directly demonstrated by the well-textured growth of an Fe/Ba--122 bilayer on MgO(001).  On bare MgO, full epitaxy was not obtained due to a large lattice misfit of around 6\%\ and $T_{\rm c}$ was significantly reduced. Second, in addition to the excellent texture of these bilayers, their $T_{\rm c}$ values remain close to that of the Ba--122 target material, thus representing a way to retain good superconducting properties in epitaxially-grown thin films.  Third, the interfacial bond between the iron layer and the Ba--122 phase is directly observed to take place on the iron sublayer within the FeAs tetrahedron using two independent imaging techniques.  Since the FeAs tetrahedron is the one structural feature common to every type of iron pnictide, these results suggest that similar bilayer structures from other iron pnictide systems can be realized.  Finally, the clean and coherent nature of the Fe/Ba--122 interface may enable the fabrication of ferromagnetic/superconducting heterostructures thus paving the way for future studies on the interplay between magnetism and superconductivity in the iron pnictides. 

\begin{acknowledgments}
We wish to acknowledge J. Scheiter for help with the TEM lamella preparation as well as S. F{\"a}hler, J. Engelmann, and S. Trommler for the scientific discussions.
\end{acknowledgments}


\begin{thebibliography}{13}%

\makeatletter
\providecommand \@ifxundefined [1]{%
 \@ifx{#1\undefined}
}%
\providecommand \@ifnum [1]{%
 \ifnum #1\expandafter \@firstoftwo
 \else \expandafter \@secondoftwo
 \fi
}%
\providecommand \@ifx [1]{%
 \ifx #1\expandafter \@firstoftwo
 \else \expandafter \@secondoftwo
 \fi
}%
\providecommand \natexlab [1]{#1}%
\providecommand \enquote  [1]{``#1''}%
\providecommand \bibnamefont  [1]{#1}%
\providecommand \bibfnamefont [1]{#1}%
\providecommand \citenamefont [1]{#1}%
\providecommand \href@noop [0]{\@secondoftwo}%
\providecommand \href [0]{\begingroup \@sanitize@url \@href}%
\providecommand \@href[1]{\@@startlink{#1}\@@href}%
\providecommand \@@href[1]{\endgroup#1\@@endlink}%
\providecommand \@sanitize@url [0]{\catcode `\\12\catcode `\$12\catcode
  `\&12\catcode `\#12\catcode `\^12\catcode `\_12\catcode `\%12\relax}%
\providecommand \@@startlink[1]{}%
\providecommand \@@endlink[0]{}%
\providecommand \url  [0]{\begingroup\@sanitize@url \@url }%
\providecommand \@url [1]{\endgroup\@href {#1}{\urlprefix }}%
\providecommand \urlprefix  [0]{URL }%
\providecommand \Eprint [0]{\href }%
\@ifxundefined \urlstyle {%
  \providecommand \doi  [0]{\begingroup \@sanitize@url \@doi}%
  \providecommand \@doi [1]{\endgroup \@@startlink {\doibase
  #1}doi:\discretionary {}{}{}#1\@@endlink }%
}{%
  \providecommand \doi  [0]{doi:\discretionary{}{}{}\begingroup
  \urlstyle{rm}\Url }%
}%
\providecommand \doibase [0]{http://dx.doi.org/}%
\providecommand \Doi [0]{\begingroup \@sanitize@url \@Doi }%
\providecommand \@Doi  [1]{\endgroup\@@startlink{\doibase#1}\@@Doi}%
\providecommand \@@Doi [1]{#1\@@endlink}%
\providecommand \selectlanguage [0]{\@gobble}%
\providecommand \bibinfo  [0]{\@secondoftwo}%
\providecommand \bibfield  [0]{\@secondoftwo}%
\providecommand \translation [1]{[#1]}%
\providecommand \BibitemOpen [0]{}%
\providecommand \bibitemStop [0]{}%
\providecommand \bibitemNoStop [0]{.\EOS\space}%
\providecommand \EOS [0]{\spacefactor3000\relax}%
\providecommand \BibitemShut  [1]{\csname bibitem#1\endcsname}%
\bibitem [{\citenamefont {Rotter}\ \emph {et~al.}(2008)\citenamefont {Rotter},
  \citenamefont {Tegel},\ and\ \citenamefont {Johrendt}}]{Rotter2008}%
  \BibitemOpen
  \bibfield  {author} {\bibinfo {author} {\bibfnamefont {M.}~\bibnamefont
  {Rotter}}, \bibinfo {author} {\bibfnamefont {M.}~\bibnamefont {Tegel}}, \
  and\ \bibinfo {author} {\bibfnamefont {D.}~\bibnamefont {Johrendt}},\ }\Doi
  {10.1103/PhysRevLett.101.107006} {\bibfield  {journal} {\bibinfo  {journal}
  {Phys. Rev. Lett.},\ }\textbf {\bibinfo {volume} {101}},\ \bibinfo {pages}
  {107006} (\bibinfo {year} {2008})}\BibitemShut {NoStop}%
\bibitem [{\citenamefont {Hiramatsu}\ \emph {et~al.}(2008)\citenamefont
  {Hiramatsu}, \citenamefont {Katase}, \citenamefont {Kamiya}, \citenamefont
  {Hirano},\ and\ \citenamefont {Hosono}}]{Hiramatsu2008}%
  \BibitemOpen
  \bibfield  {author} {\bibinfo {author} {\bibfnamefont {H.}~\bibnamefont
  {Hiramatsu}}, \bibinfo {author} {\bibfnamefont {T.}~\bibnamefont {Katase}},
  \bibinfo {author} {\bibfnamefont {T.}~\bibnamefont {Kamiya}}, \bibinfo
  {author} {\bibfnamefont {M.}~\bibnamefont {Hirano}}, \ and\ \bibinfo {author}
  {\bibfnamefont {H.}~\bibnamefont {Hosono}},\ }\Doi {10.1143/APEX.1.101702}
  {\bibfield  {journal} {\bibinfo  {journal} {Appl. Phys. Express},\ }\textbf
  {\bibinfo {volume} {1}},\ \bibinfo {pages} {101702} (\bibinfo {year}
  {2008})}\BibitemShut {NoStop}%
\bibitem [{\citenamefont {Choi}\ \emph {et~al.}(2009)\citenamefont {Choi},
  \citenamefont {Jung}, \citenamefont {Lee}, \citenamefont {Kwon},
  \citenamefont {Kang}, \citenamefont {Kim}, \citenamefont {Jung},
  \citenamefont {Lee},\ and\ \citenamefont {Sun}}]{Choi2009}%
  \BibitemOpen
  \bibfield  {author} {\bibinfo {author} {\bibfnamefont {E.-M.}\ \bibnamefont
  {Choi}}, \bibinfo {author} {\bibfnamefont {S.-G.}\ \bibnamefont {Jung}},
  \bibinfo {author} {\bibfnamefont {N.~H.}\ \bibnamefont {Lee}}, \bibinfo
  {author} {\bibfnamefont {Y.-S.}\ \bibnamefont {Kwon}}, \bibinfo {author}
  {\bibfnamefont {W.~N.}\ \bibnamefont {Kang}}, \bibinfo {author}
  {\bibfnamefont {D.~H.}\ \bibnamefont {Kim}}, \bibinfo {author} {\bibfnamefont
  {M.-H.}\ \bibnamefont {Jung}}, \bibinfo {author} {\bibfnamefont {S.-I.}\
  \bibnamefont {Lee}}, \ and\ \bibinfo {author} {\bibfnamefont
  {L.}~\bibnamefont {Sun}},\ }\Doi {10.1063/1.3204457} {\bibfield  {journal}
  {\bibinfo  {journal} {Appl. Phys. Lett.},\ }\textbf {\bibinfo {volume}
  {95}},\ \bibinfo {eid} {062507} (\bibinfo {year} {2009})}\BibitemShut
  {NoStop}%
\bibitem [{\citenamefont {Hiramatsu}\ \emph {et~al.}(2009)\citenamefont
  {Hiramatsu}, \citenamefont {Katase}, \citenamefont {Kamiya}, \citenamefont
  {Hirano},\ and\ \citenamefont {Hosono}}]{Hiramatsu2009a}%
  \BibitemOpen
  \bibfield  {author} {\bibinfo {author} {\bibfnamefont {H.}~\bibnamefont
  {Hiramatsu}}, \bibinfo {author} {\bibfnamefont {T.}~\bibnamefont {Katase}},
  \bibinfo {author} {\bibfnamefont {T.}~\bibnamefont {Kamiya}}, \bibinfo
  {author} {\bibfnamefont {M.}~\bibnamefont {Hirano}}, \ and\ \bibinfo {author}
  {\bibfnamefont {H.}~\bibnamefont {Hosono}},\ }\href@noop {} {\bibfield
  {journal} {\bibinfo  {journal} {Phys. Rev. B},\ }\textbf {\bibinfo {volume}
  {80}},\ \bibinfo {pages} {052501} (\bibinfo {year} {2009})}\BibitemShut
  {NoStop}%
\bibitem [{\citenamefont {Katase}\ \emph {et~al.}(2009)\citenamefont {Katase},
  \citenamefont {Hiramatsu}, \citenamefont {Yanagi}, \citenamefont {Kamiya},
  \citenamefont {Hirano},\ and\ \citenamefont {Hosono}}]{Katase2009a}%
  \BibitemOpen
  \bibfield  {author} {\bibinfo {author} {\bibfnamefont {T.}~\bibnamefont
  {Katase}}, \bibinfo {author} {\bibfnamefont {H.}~\bibnamefont {Hiramatsu}},
  \bibinfo {author} {\bibfnamefont {H.}~\bibnamefont {Yanagi}}, \bibinfo
  {author} {\bibfnamefont {T.}~\bibnamefont {Kamiya}}, \bibinfo {author}
  {\bibfnamefont {M.}~\bibnamefont {Hirano}}, \ and\ \bibinfo {author}
  {\bibfnamefont {H.}~\bibnamefont {Hosono}},\ }\href@noop {} {\bibfield
  {journal} {\bibinfo  {journal} {Solid State Commun.},\ }\textbf {\bibinfo
  {volume} {149}},\ \bibinfo {pages} {2121} (\bibinfo {year}
  {2009})}\BibitemShut {NoStop}%
\bibitem [{\citenamefont {Lee}\ \emph {et~al.}(2009)\citenamefont {Lee},
  \citenamefont {Jiang}, \citenamefont {Weiss}, \citenamefont {Folkman},
  \citenamefont {Bark}, \citenamefont {Tarantini}, \citenamefont {Xu},
  \citenamefont {Abraimov}, \citenamefont {Polyanskii}, \citenamefont {Nelson},
  \citenamefont {Zhang}, \citenamefont {Baek}, \citenamefont {Jang},
  \citenamefont {Yamamoto}, \citenamefont {Kametani}, \citenamefont {Pan},
  \citenamefont {Hellstrom}, \citenamefont {Gurevich}, \citenamefont {Eom},\
  and\ \citenamefont {Larbalestier}}]{Lee2009}%
  \BibitemOpen
  \bibfield  {author} {\bibinfo {author} {\bibfnamefont {S.}~\bibnamefont
  {Lee}}, \bibinfo {author} {\bibfnamefont {J.}~\bibnamefont {Jiang}}, \bibinfo
  {author} {\bibfnamefont {J.~D.}\ \bibnamefont {Weiss}}, \bibinfo {author}
  {\bibfnamefont {C.~M.}\ \bibnamefont {Folkman}}, \bibinfo {author}
  {\bibfnamefont {C.~W.}\ \bibnamefont {Bark}}, \bibinfo {author}
  {\bibfnamefont {C.}~\bibnamefont {Tarantini}}, \bibinfo {author}
  {\bibfnamefont {A.}~\bibnamefont {Xu}}, \bibinfo {author} {\bibfnamefont
  {D.}~\bibnamefont {Abraimov}}, \bibinfo {author} {\bibfnamefont
  {A.}~\bibnamefont {Polyanskii}}, \bibinfo {author} {\bibfnamefont {C.~T.}\
  \bibnamefont {Nelson}}, \bibinfo {author} {\bibfnamefont {Y.}~\bibnamefont
  {Zhang}}, \bibinfo {author} {\bibfnamefont {S.~H.}\ \bibnamefont {Baek}},
  \bibinfo {author} {\bibfnamefont {H.~W.}\ \bibnamefont {Jang}}, \bibinfo
  {author} {\bibfnamefont {A.}~\bibnamefont {Yamamoto}}, \bibinfo {author}
  {\bibfnamefont {F.}~\bibnamefont {Kametani}}, \bibinfo {author}
  {\bibfnamefont {X.~Q.}\ \bibnamefont {Pan}}, \bibinfo {author} {\bibfnamefont
  {E.~E.}\ \bibnamefont {Hellstrom}}, \bibinfo {author} {\bibfnamefont
  {A.}~\bibnamefont {Gurevich}}, \bibinfo {author} {\bibfnamefont {C.~B.}\
  \bibnamefont {Eom}}, \ and\ \bibinfo {author} {\bibfnamefont {D.~C.}\
  \bibnamefont {Larbalestier}},\ }\Doi {10.1063/1.3262953} {\bibfield
  {journal} {\bibinfo  {journal} {Appl. Phys. Lett.},\ }\textbf {\bibinfo
  {volume} {95}},\ \bibinfo {eid} {212505} (\bibinfo {year}
  {2009})}\BibitemShut {NoStop}%
\bibitem [{\citenamefont {Iida}\ \emph {et~al.}(2009)\citenamefont {Iida},
  \citenamefont {Hanisch}, \citenamefont {Huhne}, \citenamefont {Kurth},
  \citenamefont {Kidszun}, \citenamefont {Haindl}, \citenamefont {Werner},
  \citenamefont {Schultz},\ and\ \citenamefont {Holzapfel}}]{Iida2009}%
  \BibitemOpen
  \bibfield  {author} {\bibinfo {author} {\bibfnamefont {K.}~\bibnamefont
  {Iida}}, \bibinfo {author} {\bibfnamefont {J.}~\bibnamefont {Hanisch}},
  \bibinfo {author} {\bibfnamefont {R.}~\bibnamefont {Huhne}}, \bibinfo
  {author} {\bibfnamefont {F.}~\bibnamefont {Kurth}}, \bibinfo {author}
  {\bibfnamefont {M.}~\bibnamefont {Kidszun}}, \bibinfo {author} {\bibfnamefont
  {S.}~\bibnamefont {Haindl}}, \bibinfo {author} {\bibfnamefont
  {J.}~\bibnamefont {Werner}}, \bibinfo {author} {\bibfnamefont
  {L.}~\bibnamefont {Schultz}}, \ and\ \bibinfo {author} {\bibfnamefont
  {B.}~\bibnamefont {Holzapfel}},\ }\Doi {10.1063/1.3259922} {\bibfield
  {journal} {\bibinfo  {journal} {Appl. Phys. Lett.},\ }\textbf {\bibinfo
  {volume} {95}},\ \bibinfo {eid} {192501} (\bibinfo {year}
  {2009})}\BibitemShut {NoStop}%
\bibitem [{\citenamefont {Lee}\ \emph {et~al.}(2010)\citenamefont {Lee},
  \citenamefont {Jiang}, \citenamefont {Zhang}, \citenamefont {Bark},
  \citenamefont {Weiss}, \citenamefont {Tarantini}, \citenamefont {Nelson},
  \citenamefont {Jang}, \citenamefont {Folkman}, \citenamefont {Baek},
  \citenamefont {Polyanskii}, \citenamefont {Abraimov}, \citenamefont
  {Yamamoto}, \citenamefont {Park}, \citenamefont {Pan}, \citenamefont
  {Hellstrom}, \citenamefont {Larbalestier},\ and\ \citenamefont
  {Eom}}]{Lee2010}%
  \BibitemOpen
  \bibfield  {author} {\bibinfo {author} {\bibfnamefont {S.}~\bibnamefont
  {Lee}}, \bibinfo {author} {\bibfnamefont {J.}~\bibnamefont {Jiang}}, \bibinfo
  {author} {\bibfnamefont {Y.}~\bibnamefont {Zhang}}, \bibinfo {author}
  {\bibfnamefont {C.~W.}\ \bibnamefont {Bark}}, \bibinfo {author}
  {\bibfnamefont {J.~D.}\ \bibnamefont {Weiss}}, \bibinfo {author}
  {\bibfnamefont {C.}~\bibnamefont {Tarantini}}, \bibinfo {author}
  {\bibfnamefont {C.~T.}\ \bibnamefont {Nelson}}, \bibinfo {author}
  {\bibfnamefont {H.~W.}\ \bibnamefont {Jang}}, \bibinfo {author}
  {\bibfnamefont {C.~M.}\ \bibnamefont {Folkman}}, \bibinfo {author}
  {\bibfnamefont {S.~H.}\ \bibnamefont {Baek}}, \bibinfo {author}
  {\bibfnamefont {A.}~\bibnamefont {Polyanskii}}, \bibinfo {author}
  {\bibfnamefont {D.}~\bibnamefont {Abraimov}}, \bibinfo {author}
  {\bibfnamefont {A.}~\bibnamefont {Yamamoto}}, \bibinfo {author}
  {\bibfnamefont {J.~W.}\ \bibnamefont {Park}}, \bibinfo {author}
  {\bibfnamefont {X.~Q.}\ \bibnamefont {Pan}}, \bibinfo {author} {\bibfnamefont
  {E.~E.}\ \bibnamefont {Hellstrom}}, \bibinfo {author} {\bibfnamefont {D.~C.}\
  \bibnamefont {Larbalestier}}, \ and\ \bibinfo {author} {\bibfnamefont
  {C.~B.}\ \bibnamefont {Eom}},\ }\href {http://dx.doi.org/10.1038/nmat2721}
  {\bibfield  {journal} {\bibinfo  {journal} {Nat. Mater.},\ }\textbf {\bibinfo
  {volume} {9}},\ \bibinfo {pages} {397 -- 402} (\bibinfo {year} {2010})},\
  ISSN \bibinfo {issn} {1476-4660}\BibitemShut {NoStop}%
\bibitem [{\citenamefont {Iida}\ \emph {et~al.}(2010)\citenamefont {Iida},
  \citenamefont {H{\"a}nisch}, \citenamefont {Thersleff}, \citenamefont
  {Kurth}, \citenamefont {Kidszun}, \citenamefont {Haindl}, \citenamefont
  {H{\"u}hne}, \citenamefont {Schultz},\ and\ \citenamefont
  {Holzapfel}}]{Iida2010a}%
  \BibitemOpen
  \bibfield  {author} {\bibinfo {author} {\bibfnamefont {K.}~\bibnamefont
  {Iida}}, \bibinfo {author} {\bibfnamefont {J.}~\bibnamefont {H{\"a}nisch}},
  \bibinfo {author} {\bibfnamefont {T.}~\bibnamefont {Thersleff}}, \bibinfo
  {author} {\bibfnamefont {F.}~\bibnamefont {Kurth}}, \bibinfo {author}
  {\bibfnamefont {M.}~\bibnamefont {Kidszun}}, \bibinfo {author} {\bibfnamefont
  {S.}~\bibnamefont {Haindl}}, \bibinfo {author} {\bibfnamefont
  {R.}~\bibnamefont {H{\"u}hne}}, \bibinfo {author} {\bibfnamefont
  {L.}~\bibnamefont {Schultz}}, \ and\ \bibinfo {author} {\bibfnamefont
  {B.}~\bibnamefont {Holzapfel}},\ }\href@noop {} {\bibfield  {journal}
  {\bibinfo  {journal} {Phys. Rev. B},\ }\textbf {\bibinfo {volume} {81}},\
  \bibinfo {pages} {100507(R)} (\bibinfo {year} {2010})}\BibitemShut {NoStop}%
\bibitem [{\citenamefont {Katase}\ \emph {et~al.}(2010)\citenamefont {Katase},
  \citenamefont {Ishimaru}, \citenamefont {Tsukamoto}, \citenamefont
  {Hiramatsu}, \citenamefont {Kamiya}, \citenamefont {Tanabe},\ and\
  \citenamefont {Hosono}}]{Katase2010}%
  \BibitemOpen
  \bibfield  {author} {\bibinfo {author} {\bibfnamefont {T.}~\bibnamefont
  {Katase}}, \bibinfo {author} {\bibfnamefont {Y.}~\bibnamefont {Ishimaru}},
  \bibinfo {author} {\bibfnamefont {A.}~\bibnamefont {Tsukamoto}}, \bibinfo
  {author} {\bibfnamefont {H.}~\bibnamefont {Hiramatsu}}, \bibinfo {author}
  {\bibfnamefont {T.}~\bibnamefont {Kamiya}}, \bibinfo {author} {\bibfnamefont
  {K.}~\bibnamefont {Tanabe}}, \ and\ \bibinfo {author} {\bibfnamefont
  {H.}~\bibnamefont {Hosono}},\ }\href {arXiv:condmat/1001.3615} {\bibfield
  {journal} {\bibinfo  {journal} {Appl. Phys. Lett.},\ }\textbf {\bibinfo
  {volume} {96}},\ \bibinfo {pages} {142507} (\bibinfo {year}
  {2010})}\BibitemShut {NoStop}%
\bibitem [{\citenamefont {Jordan}\ \emph {et~al.}(1998)\citenamefont {Jordan},
  \citenamefont {Lawler}, \citenamefont {Schad},\ and\ \citenamefont {van
  Kempen}}]{Jordan1998}%
  \BibitemOpen
  \bibfield  {author} {\bibinfo {author} {\bibfnamefont {S.~M.}\ \bibnamefont
  {Jordan}}, \bibinfo {author} {\bibfnamefont {J.~F.}\ \bibnamefont {Lawler}},
  \bibinfo {author} {\bibfnamefont {R.}~\bibnamefont {Schad}}, \ and\ \bibinfo
  {author} {\bibfnamefont {H.}~\bibnamefont {van Kempen}},\ }\href@noop {}
  {\bibfield  {journal} {\bibinfo  {journal} {J. Appl. Phys.},\ }\textbf
  {\bibinfo {volume} {84}},\ \bibinfo {pages} {1499--1503} (\bibinfo {year}
  {1998})}\BibitemShut {NoStop}%
\bibitem [{\citenamefont {Langford}(2006)}]{Langford06}%
  \BibitemOpen
  \bibfield  {author} {\bibinfo {author} {\bibfnamefont {R.~M.}\ \bibnamefont
  {Langford}},\ }\Doi {10.1002/jemt.20324} {\bibfield  {journal} {\bibinfo
  {journal} {Microsc. Res. Techniq.},\ }\textbf {\bibinfo {volume} {69}},\
  \bibinfo {pages} {538--549} (\bibinfo {year} {2006})}\BibitemShut {NoStop}%
\end{thebibliography}
%

\end{document}